\input amssym.tex

% Page layout

\magnification=\magstephalf
\hsize=14.0 true cm
\vsize=19 true cm
\hoffset=1.0 true cm
\voffset=2.0 true cm

\abovedisplayskip=12pt plus 3pt minus 3pt
\belowdisplayskip=12pt plus 3pt minus 3pt
\parindent=1.0em

% Fonts

\font\sixrm=cmr6
\font\eightrm=cmr8
\font\ninerm=cmr9

\font\sixi=cmmi6
\font\eighti=cmmi8
\font\ninei=cmmi9

\font\sixsy=cmsy6
\font\eightsy=cmsy8
\font\ninesy=cmsy9

\font\sixbf=cmbx6
\font\eightbf=cmbx8
\font\ninebf=cmbx9

\font\eightit=cmti8
\font\nineit=cmti9

\font\eightsl=cmsl8
\font\ninesl=cmsl9

\font\sixss=cmss8 at 8 true pt
\font\sevenss=cmss9 at 9 true pt
\font\eightss=cmss8
\font\niness=cmss9
\font\tenss=cmss10

 at 12 true pt
 at 12 true pt
\font\bigrm=cmr10 at 12 true pt
 at 12 true pt
 at 12 true pt

 at 16 true pt
%\font\Bigsy=cmsy12 at 16 true pt
%\font\Bigex=cmex12 at 16 true pt
 at 16 true pt
\font\Bigrm=cmr12 at 16 true pt
 at 16 true pt
 at 16 true pt

\catcode`@=11
\newfam\ssfam

\def\tenpoint{\def\rm{\fam0\tenrm}%
    \textfont0=\tenrm \scriptfont0=\sevenrm \scriptscriptfont0=\fiverm
    \textfont1=\teni  \scriptfont1=\seveni  \scriptscriptfont1=\fivei
    \textfont2=\tensy \scriptfont2=\sevensy \scriptscriptfont2=\fivesy
    \textfont3=\tenex \scriptfont3=\tenex   \scriptscriptfont3=\tenex
    \textfont\itfam=\tenit                  \def\it{\fam\itfam\tenit}%
    \textfont\slfam=\tensl                  \def\sl{\fam\slfam\tensl}%
    \textfont\bffam=\tenbf \scriptfont\bffam=\sevenbf
    \scriptscriptfont\bffam=\fivebf
                                            \def\bf{\fam\bffam\tenbf}%
    \textfont\ssfam=\tenss \scriptfont\ssfam=\sevenss
    \scriptscriptfont\ssfam=\sevenss
                                            \def\ss{\fam\ssfam\tenss}%
    \normalbaselineskip=13pt
    \setbox\strutbox=\hbox{\vrule height8.5pt depth3.5pt width0pt}%
    \let\big=\tenbig
    \normalbaselines\rm}

\def\ninepoint{\def\rm{\fam0\ninerm}%
    \textfont0=\ninerm      \scriptfont0=\sixrm
                            \scriptscriptfont0=\fiverm
    \textfont1=\ninei       \scriptfont1=\sixi
                            \scriptscriptfont1=\fivei
    \textfont2=\ninesy      \scriptfont2=\sixsy
                            \scriptscriptfont2=\fivesy
    \textfont3=\tenex       \scriptfont3=\tenex
                            \scriptscriptfont3=\tenex
    \textfont\itfam=\nineit \def\it{\fam\itfam\nineit}%
    \textfont\slfam=\ninesl \def\sl{\fam\slfam\ninesl}%
    \textfont\bffam=\ninebf \scriptfont\bffam=\sixbf
                            \scriptscriptfont\bffam=\fivebf
                            \def\bf{\fam\bffam\ninebf}%
    \textfont\ssfam=\niness \scriptfont\ssfam=\sixss
                            \scriptscriptfont\ssfam=\sixss
                            \def\ss{\fam\ssfam\niness}%
    \normalbaselineskip=12pt
    \setbox\strutbox=\hbox{\vrule height8.0pt depth3.0pt width0pt}%
    \let\big=\ninebig
    \normalbaselines\rm}

\def\eightpoint{\def\rm{\fam0\eightrm}%
    \textfont0=\eightrm      \scriptfont0=\sixrm
                             \scriptscriptfont0=\fiverm
    \textfont1=\eighti       \scriptfont1=\sixi
                             \scriptscriptfont1=\fivei
    \textfont2=\eightsy      \scriptfont2=\sixsy
                             \scriptscriptfont2=\fivesy
    \textfont3=\tenex        \scriptfont3=\tenex
                             \scriptscriptfont3=\tenex
    \textfont\itfam=\eightit \def\it{\fam\itfam\eightit}%
    \textfont\slfam=\eightsl \def\sl{\fam\slfam\eightsl}%
    \textfont\bffam=\eightbf \scriptfont\bffam=\sixbf
                             \scriptscriptfont\bffam=\fivebf
                             \def\bf{\fam\bffam\eightbf}%
    \textfont\ssfam=\eightss \scriptfont\ssfam=\sixss
                             \scriptscriptfont\ssfam=\sixss
                             \def\ss{\fam\ssfam\eightss}%
    \normalbaselineskip=10pt
    \setbox\strutbox=\hbox{\vrule height7.0pt depth2.0pt width0pt}%
    \let\big=\eightbig
    \normalbaselines\rm}

\def\tenbig#1{{\hbox{$\left#1\vbox to8.5pt{}\right.\n@space$}}}
\def\ninebig#1{{\hbox{$\textfont0=\tenrm\textfont2=\tensy
                       \left#1\vbox to7.25pt{}\right.\n@space$}}}
\def\eightbig#1{{\hbox{$\textfont0=\ninerm\textfont2=\ninesy
                       \left#1\vbox to6.5pt{}\right.\n@space$}}}

\font\sectionfont=cmbx10
\font\subsectionfont=cmti10

\def\figurecaptionfont{\ninepoint}
\def\tablecaptionfont{\ninepoint}
\def\footnotefont{\eightpoint}

% New count registers

\newcount\equationno
\newcount\bibitemno
\newcount\figureno
\newcount\tableno

\equationno=0
\bibitemno=0
\figureno=0
\tableno=0
%\advance\pageno by -1

% Footline

\footline={\ifnum\pageno=0{\hfil}\else
{\hss\rm\the\pageno\hss}\fi}

% Section macro

\def\section #1 \par
{\vskip3ex
\global\def\equationlabel{#1}
\global\equationno=0
\immediate\write\terminal{Section #1.}}

% Subsection macro

\def\subsection #1 \par
{\vskip0pt plus 0.8 true cm\penalty-50 \vskip0pt plus-0.8 true cm
\vskip2.5ex plus 0.1ex minus 0.1ex
\leftline{\subsectionfont #1}\par
\immediate\write\terminal{Subsection #1}
\vskip1.0ex plus 0.1ex minus 0.1ex
\noindent}

% Appendix macro

\def\appendix #1. #2 \par
{\vskip0pt plus .20\vsize\penalty-100 \vskip0pt plus-.20\vsize
\vskip 1.6 true cm plus 0.2 true cm minus 0.2 true cm
\global\def\equationlabel{\hbox{\rm#1}}
\global\equationno=0
\leftline{\sectionfont Appendix #1. #2}\par
\immediate\write\terminal{Appendix #1. #2}
\vskip 0.7 true cm plus 0.1 true cm minus 0.1 true cm
\noindent}

%\def\appendix #1. #2 \par
%{\vskip0pt plus .20\vsize\penalty-100 \vskip0pt plus-.20\vsize
%\vskip 1.6 true cm plus 0.2 true cm minus 0.2 true cm
%\global\def\equationlabel{\hbox{\rm#1}}
%\global\equationno=0
%\leftline{\sectionfont Appendix #1. #2}\par
%\immediate\write\terminal{Appendix #1. #2}
%\vskip 0.7 true cm plus 0.1 true cm minus 0.1 true cm
%\noindent}

% Displayed equations

\def\equation#1{$$\displaylines{\qquad #1}$$}
\def\enum{\global\advance\equationno by 1
\hfill\llap{{\rm(\equationlabel.\the\equationno)}}}
\def\noenum{\hfill}
\def\next#1{\cr\noalign{\vskip#1}\qquad}

% Bibliography macro, references

\def\ifundefined#1{\expandafter\ifx\csname#1\endcsname\relax}

\def\ref#1{\ifundefined{#1}?\immediate\write\terminal{unknown reference
on page \the\pageno}\else\csname#1\endcsname\fi}

\newwrite\terminal
\newwrite\bibitemlist

\def\bibitem#1#2\par{\global\advance\bibitemno by 1
\immediate\write\bibitemlist{\string\def
\expandafter\string\csname#1\endcsname
{\the\bibitemno}}
\item{[\the\bibitemno]}#2\par}

\def\beginbibliography{
\vskip0pt plus .15\vsize\penalty-100 \vskip0pt plus-.15\vsize
\vskip 1.2 true cm plus 0.2 true cm minus 0.2 true cm
\centerline{\sectionfont References}\par
\immediate\write\terminal{References}
\immediate\openout\bibitemlist=biblist
\frenchspacing\parindent=1.8em
\vskip 0.5 true cm plus 0.1 true cm minus 0.1 true cm}

\def\endbibliography{
\immediate\closeout\bibitemlist
\nonfrenchspacing\parindent=1.0em}

% Figure and table captions

\def\figurecaption#1{\global\advance\figureno by 1
\narrower\figurecaptionfont
Fig.~\the\figureno. #1}

\def\tablecaption#1{\global\advance\tableno by 1
\vbox to 0.5 true cm { }
\centerline{\tablecaptionfont%
Table~\the\tableno. #1}
\vskip-0.4 true cm}

\tenpoint

\immediate\openin\bibitemlist=biblist
\ifeof\bibitemlist\immediate\closein\bibitemlist
\else\immediate\closein\bibitemlist
\input biblist \fi

% current year and month

\def\thismonth{\ifcase\month\or
January\or February\or March\or April\or May\or June\or
July\or August\or September\or October\or November\or December\fi}

% Definitions and abbreviations

% Roman letters in math formulae

\def\rmd{{\rm d}}

\def\rme{{\rm e}}

\def\rmU{{\rm U}}

% Real and integer numbers

\def\rz{{\Bbb R}}

% Special relations and symbols

\def\proof{\noindent{\sl Proof:}\kern0.6em}

\def\frac#1#2{\hbox{$#1\over#2$}}
\def\dual{\mathstrut^*\kern-0.1em}

\def\lvec#1{\setbox0=\hbox{$#1$}
    \setbox1=\hbox{$\scriptstyle\leftarrow$}
    #1\kern-\wd0\smash{
    \raise\ht0\hbox{$\raise1pt\hbox{$\scriptstyle\leftarrow$}$}}
    \kern-\wd1\kern\wd0}
\def\rvec#1{\setbox0=\hbox{$#1$}
    \setbox1=\hbox{$\scriptstyle\rightarrow$}
    #1\kern-\wd0\smash{
    \raise\ht0\hbox{$\raise1pt\hbox{$\scriptstyle\rightarrow$}$}}
    \kern-\wd1\kern\wd0}
\def\slash#1{\setbox0=\hbox{$#1$}\setbox1=\hbox{$\kern1pt/$}
    #1\kern-\wd0\kern1pt/\kern-\wd1\kern\wd0}

% Lattice derivatives

\def\nabstar#1{{\nabla\kern0.5pt\smash{\raise 4.5pt\hbox{$\ast$}}
               \kern-5.5pt_{#1}}}

\def\drvstar#1{{\partial\kern0.5pt\smash{\raise 4.5pt\hbox{$\ast$}}
               \kern-6.0pt_{#1}}}

\def\ldrvstar#1{{\lvec{\,\partial}\kern-0.5pt\smash{\raise 4.5pt\hbox{$\ast$}}
               \kern-5.0pt_{#1}}}

% Units

% Constants

% Fields

\def\psibar{\overline{\psi}}

% Dirac matrices

\def\dirac#1{\gamma_{#1}}
\def\diracstar#1#2{
    \setbox0=\hbox{$\gamma$}\setbox1=\hbox{$\gamma_{#1}$}
    \gamma_{#1}\kern-\wd1\kern\wd0
    \smash{\raise4.5pt\hbox{$\scriptstyle#2$}}}

% Gauge group

\def\tr{{\rm tr}}
\def\Tr{{\rm Tr}}
\def\Ad{{\rm Ad}\kern0.1em}

% Dirac operator

% Quark masses

\def\Nf{N_{\rm f}}

%Connected correlation functions

\def\con{_{\raise0.9pt\hbox{\kern-1pt\sevenrm c}}}%
\vbox{\vskip0.0cm}
\rightline{CERN-PH-TH/2004-062}

\vskip 2.0cm 
\centerline{\Bigrm Topological effects in QCD and the problem of}
\vskip 0.3 true cm
\centerline{\Bigrm short-distance singularities}
\vskip 0.6 true cm
\centerline{\bigrm Martin L\"uscher}
\vskip1ex
\centerline{\it CERN, Physics Department, TH Division}
\centerline{\it CH-1211 Geneva 23, Switzerland}

\vskip 2.0 true cm
\ninepoint
\centerline{\bf Abstract}
\vskip 2.0ex
The topological susceptibility and 
the higher moments of the topological charge distribution
in QCD are expressed through
certain $n$-point functions of the 
scalar and pseudo-scalar quark densities at vanishing momenta, 
which are free of
short-distance singularities. Since the normalization 
of the correlation functions is determined
by the non-singlet chiral Ward identities, these formulae provide 
an unambiguous regularization-independent definition of the moments
and thus of the charge distribution.
\tenpoint

\vfill\eject

\section 1

{\bf 1.}~Quantum fields are subject to random fluctuations
and are therefore not an obvious case for the application of 
topological arguments.
In the presence of a smooth 
background field,
and if the quantum fluctuations are treated perturbatively,
the problem usually remains unnoticed, because 
the topological information is entirely encoded in the background field. 
In general a separation of the fluctuations is not possible, however, 
and in these situations it may be difficult to give an unambiguous
meaning to the notion of, say, a topological sector.

In QCD the axial anomaly provides a link between 
the correlation functions of local fields and the topology of 
the underlying classical field space.
Large-$N$ 
counting rules and
the anomalous chiral Ward identities lead to 
the Witten--Veneziano formula [\ref{Witten}--\ref{Seiler}],
for example, which relates the vacuum expectation value of the 
square of the topological charge
\equation{
  Q=\int\rmd^4x\, q(x),
  \qquad q(x)\equiv-{1\over32\pi^2}\,\epsilon_{\mu\nu\rho\sigma}
  \tr\{F_{\mu\nu}(x)F_{\rho\sigma}(x)\},
  \enum
}
(where $F_{\mu\nu}$ denotes the field strength of the
gauge field) to the mass of the $\eta'$ meson.
More recently,  
when studying QCD in the so-called $\epsilon$-regime
[\ref{GasserLeutwyler}],
the Ward identity
was used to trade the topological term in the partition function
\equation{
  {\cal Z}(\theta)=\int_{\rm fields}\rme^{-S+i\theta Q}
  \enum
}
for a phase factor in the quark mass term in the action $S$.
Close to the chiral limit, and at any value of $\theta$, 
the theory may then be described by the standard effective chiral 
$\sigma$-model [\ref{LeutwylerSmilga}].
An interesting point to note here is that, in the effective
theory, it is possible to define fixed-charge correlation
functions by keeping only the terms proportional to a given power 
of $\rme^{i\theta}$ in the chiral expansions of the partition function
and the unnormalized correlation functions.

For the reasons indicated above, the situation in QCD is
more complicated in this respect. 
Technically the difficulty is that
the topological susceptibility
\equation{
   \chi_t=\int\rmd^4x\,\langle q(x)q(0)\rangle
   \enum
}
and the higher derivatives of the
free energy $F(\theta)=-\ln{\cal Z}(\theta)$
involve an integration of correlation functions with non-integrable
short-distance singularities.
In eq.~(1.3), for example,
the integrand 
diverges like $(x^2)^{-4}$ (up to logarithms).
If no subtraction prescription is specified, the free energy
is therefore ill-defined, and so are
the fixed-charge correlation functions of local fields.

It is sometimes pointed out in this connection that the 
charge density is equal to the divergence of the Chern--Simons current
and that this would allow the integral to be rewritten
in the form of a surface integral. While 
the short-distance singularities are avoided in this way,
the argument requires the gauge to be fixed and is
hence difficult to put on solid grounds
beyond perturbation theory.

\section 2

{\bf 2.}~In lattice QCD various
definitions of the topological susceptibility were proposed
over the years, but
it remained unclear which of these (if any) would 
have the correct continuum limit. The question is 
in fact undecidable as long as there is no
unambiguous definition of the susceptibility
in the continuum theory.

The charge density may be represented on the lattice
by any local composite field of dimension $4$
with the appropriate symmetry properties.
In general the topological character of the charge density
is lost, however, and
the topological susceptibility
consequently does not vanish in perturbation theory
(and therefore diverges proportionally to the fourth inverse power
of the lattice spacing) [\ref{DiVecchiaEtAl}].
To avoid this singular behaviour
one would like the charge density on the lattice to be such
that the associated charge is 
invariant under smooth deformations 
of the gauge field.
In the classical continuum theory,
the density is, incidentally, completely determined by
this condition up to
divergence terms $\partial_{\mu}k_{\mu}$
(where $k_{\mu}$ is any gauge-invariant local current)
and a normalization factor
[\ref{BrandtEtAl},\ref{DuboisVioletteEtAl}].

Lattice representations of the charge density 
that preserve its topological character
were constructed a long time ago
[\ref{TopA},\ref{TopB}].
Unfortunately these constructions are not unique,
because the space of lattice gauge fields is connected
and its division into charge sectors is therefore 
arbitrary to some extent.
Since there is an action barrier between the sectors,
it seems likely, however, that the regions in field space
around the sector boundaries become irrelevant close to the 
continuum limit.
Whether the topological susceptibility is finite
in this limit 
is then a second issue that remains to be discussed.

If one of the
formulations of lattice QCD that preserve chiral symmetry
is used,
there is a natural definition of the charge density
that is topological in the sense explained above and that
appears in the flavour-singlet chiral Ward identities,
exactly as in the continuum theory  [\ref{GinspargWilson}--\ref{BoulderReview}].
For this case,
and if there are $3$ or more mass-degenerate quark flavours,
Giusti, Rossi and Testa [\ref{GiustiRossiTesta}] recently showed that the 
topological susceptibility is at most logarithmically 
divergent in the continuum limit, i.e.~that 
all power-divergent contributions cancel. 

In the present paper the argumentation of ref.~[\ref{GiustiRossiTesta}] 
is cast in a slightly different and more general form.
This leads to an alternative representation of the susceptibility
and the higher-order moments of the charge distribution,
which remains well-defined in the continuum limit.
The lattice regularization is then no longer needed
and the final formulae (which are free of 
short-distance singularities) may be taken 
as the field-theoretic definition of the susceptibility and the 
higher moments.

\section 3

{\bf 3.}~In the following 
we consider QCD with
any number $\Nf$ of massive quarks.
Space-time is assumed to be a finite periodic box,
but in many equations it will be straightforward 
to pass to the infinite-volume limit
since the theory has a mass gap.
We now first derive a new representation of the topological
susceptibility on the lattice 
and shall return to the continuum theory in the next section.

The lattice theory is set up 
on a hypercubic lattice with spacing $a$,
using a (massless) lattice 
Dirac operator $D$ that satisfies the Ginsparg--Wilson relation
\equation{
  \dirac{5}D+D\dirac{5}=a D\dirac{5}D
  \enum
}
and the usual symmetry, hermiticity and regularity 
requirements \footnote{$\dagger$}{\footnotefont
The Dirac matrices $\dirac{\mu}$ are taken to be hermitian and
$\dirac{5}=-\dirac{0}\dirac{1}\dirac{2}\dirac{3}$.
Whenever possible
the same symbols are used for the fields (and other items)
in the continuum theory and 
on the lattice since it is usually clear from the context which
one is meant.}.
As explained in ref.~[\ref{BoulderReview}], the 
suggested choice of the quark action in this formulation of lattice QCD 
reads
\equation{
  S_{\rm F}=a^4\sum_{x,r}\psibar_r(x) 
  D_{m_r}\psi_r(x),
  \qquad
  D_m\equiv (1-\frac{1}{2}a m)D+m,
  \enum
}
where the index $r$ labels the quark flavours and
$m_1,\ldots,m_{\Nf}$ are the associated bare quark masses.
Most other details of the lattice theory (the action of the 
gauge field for example) are left unspecified since we 
will not need to know them.

The scalar and pseudo-scalar quark densities
that transform like a $\rmU(\Nf)\times\rmU(\Nf)$ 
multiplet under the exact chiral symmetries
of the lattice action (3.2) are
given by
\equation{
   S_{rs}(x)=\psibar_r(x)(1-\frac{1}{2}a D)\psi_s(x),
   \enum
   \next{2ex}
   P_{rs}(x)=\psibar_r(x)\dirac{5}
   (1-\frac{1}{2}a D)\psi_s(x).
   \enum
}
Keeping track of the flavour indices in this way 
rather than introducing a basis of group generators will simplify
the discussion in the following.

As is the case in the continuum theory, 
the flavour-singlet chiral transformations are anomalous on the lattice,
the anomaly being proportional to 
the topological charge density 
\equation{
   q(x)=-\frac{1}{2}a\kern1pt\tr\{\dirac{5}D(x,x)\}.
   \enum
}
In this equation,
$D(x,y)$ stands for the kernel of the lattice
Dirac operator in position space and the little trace ``tr" is taken 
over the Dirac and 
colour indices only.
The normalizations are such that 
the associated charge 
\equation{
   Q=a^4\sum_x q(x)
   \enum
}
is equal to the index of the Dirac operator
[\ref{HLN}].

A well-known consequence of 
the Ginsparg--Wilson relation
and the $\dirac{5}$-hermiticity of the Dirac operator 
is that, for any given gauge field, there exists an
orthonormal basis of eigenfunctions
of $D$ with eigenvalues $\lambda$ of the form
\equation{
   \lambda={1\over a}\left(1-\rme^{i\alpha}\right),
   \quad \alpha\in\rz.
   \enum
}
Now if $f(\lambda)$ is any
bounded function on the spectral circle (3.7), 
the identity
\equation{
   \Tr\kern-1pt\left\{\dirac{5}f(D)\right\}=\left\{f(0)-f(2/a)\right\}Q
   \enum
}
is easily established by evaluating the trace in the basis 
of eigenfunctions $\chi$ of $D$, 
noting that $\dirac{5}\chi$ is orthogonal to $\chi$ 
if the associated eigenvalue $\lambda$ is not real.

Using this general result, 
a variety of formulae
for the topological charge can be obtained.
We may, for example, start from the equation
\equation{
   a^{4r}\sum_{x_1,\ldots,x_r}
   \left\langle P_{r1}(x_1)S_{12}(x_2)\ldots S_{r-1\kern1pt r}(x_r)\right
   \rangle_{\rm F}=
   \noenum
   \next{1.5ex}
   \hskip7.0em 
   -\Tr\!\left\{\dirac{5}(1-\frac{1}{2}aD)(D_{m_1})^{-1}\ldots
                 (1-\frac{1}{2}aD)(D_{m_r})^{-1}\right\}
   \enum
} 
in which 
$\left\langle\ldots\right\rangle_{\rm F}$ denotes the fermion expectation value
(the Wick contraction)
of the fields in the bracket.
The application of the trace identity (3.8) then yields
\equation{
   Q=-m_1\ldots m_r\times
   a^{4r}\sum_{x_1,\ldots,x_r}
   \left\langle P_{r1}(x_1)S_{12}(x_2)\ldots S_{r-1\kern1pt r}(x_r)
   \right\rangle_{\rm F}
   \enum
}
for all $r\leq\Nf$.

It is now immediate that 
\equation{
   \chi_t=
   m_1\ldots m_s
   \times a^{4s-4}\sum_{x_1,\ldots,x_{s-1}}
   \bigl\langle
   P_{r1}(x_1)S_{12}(x_2)\ldots S_{r-1\kern1pt r}(x_{r})
   \noenum
   \next{1.5ex}
   \hskip6.7em
   \times P_{s\kern1pt r+1}(x_{r+1})S_{r+1\kern1pt r+2}(x_{r+2})\ldots 
   S_{s-1\kern1pt s}(0)
   \bigr\rangle\con
  \enum
}
if $r,s$ are in the range $1\leq r<s\leq\Nf$. 
In this formula, 
$\langle\ldots\rangle_{\raise-0.5pt\hbox{\sevenrm\kern-0.3pt c}}$ 
denotes the 
full QCD connected correlation function of the local fields in the bracket
and the flavour labels are such that the contraction
of the quark fields results in a product of two fermion loops
(one for each factor of the topological charge).

\section 4

{\bf 4.}~Assuming for a moment that there are at least $5$ quark flavours,
we are thus led to tentatively write 
\equation{
   \chi_t=
   m_1\ldots m_5
   \int\rmd^4x_1\ldots\rmd^4x_4
   \bigl\langle
   P_{31}(x_1)S_{12}(x_2)S_{23}(x_3)P_{54}(x_4)S_{45}(0)
   \bigr\rangle\con
  \enum
}
for the topological susceptibility in the continuum theory.
Power counting and 
the operator product expansion now suggest that the correlation function
in this expression does not have any non-integrable
short-distance singularities. 
The dimensionality of the correlation function 
certainly excludes such a singularity if all coordinates
are scaled to a common point.
If only some coordinates are scaled,
the corresponding field product converges to a local
field of dimension $3$ or more, 
and the dimensional analysis then again shows that
there is no non-integrable singularity.

A second observation is that the normalization of 
the products $m_rP_{st}$ and $m_rS_{st}$ is determined
by the non-singlet chiral Ward identities.
Equation (4.1) thus provides an
unambiguous regularization-independent definition
of the topological susceptibility in the continuum theory 
(if there are $5$ or more flavours of quarks).
Moreover we may conclude that 
the susceptibility on the lattice, as
defined in section 3,
is finite in the continuum limit
and that its value in the limit is given by eq.~(4.1).

It may be reassuring to note at this point 
that trace identities similar to those 
we have used on the lattice
hold in the continuum theory too.
In the presence of a smooth background gauge field,
the massless Dirac operator $D$
has all its eigenvalues $\lambda$ on the imaginary
axis in this case, and if $f(\lambda)$ is any continuous
function that decays rapidly enough at infinity,
it can be shown that
\equation{
  \Tr\kern-1pt\left\{\dirac{5}f(D)\right\}=f(0)Q.
  \enum
}
A relatively easy proof of the Atiyah--Singer index theorem actually starts
from the observation that
\equation{
  \Tr\bigl\{\dirac{5}\rme^{tD^2}\bigr\}=\hbox{index}(D)
  \enum
}
for any $t>0$. 
In the limit $t\to0$,
the trace can then be worked out,
using heat kernel techniques,
and is found to 
be equal to the topological charge of the background field 
[\ref{Gilkey}].
The application of eq.~(4.1) in a semi-classical context 
is therefore guaranteed to give results consistent with naive expectations.

\section 5

{\bf 5.}~The
higher connected moments
\equation{
  C_n=a^{8n-4}\sum_{x_1,\ldots,x_{2n-1}}
  \bigl\langle q(x_1)\ldots q(x_{2n-1})q(0)\bigr\rangle\con
  \enum
}
of the charge distribution on the lattice
can also be rewritten in a form that remains
well-defined in the continuum limit.
Assuming again that there are at least $5$ quark flavours,
we have
\equation{
  C_n=m_1\ldots m_5\times
  a^{8n+8}\sum_{x_1,\ldots,x_{2n+2}}
  \bigl\langle
  P_{31}(x_1)S_{12}(x_2)S_{23}(x_3)P_{54}(x_4)S_{45}(x_5)
  \noenum
  \next{0.5ex}
  \hskip15.3em
  \times q(x_6)\ldots q(x_{2n+2})q(0)
  \bigr\rangle\con.
  \enum
}
This formula is not quite what we need (some sub-leading
non-integrable singularities are still present),
but we may now eliminate
the remaining factors of the topological density
by applying
the exact chiral Ward identity [\ref{ExactChSy}]
\equation{
  a^4\sum_x\bigl\langle\left\{q(x)+m_1P_{11}(x)\right\}{\cal O}\bigr\rangle\con=
  \frac{1}{2}\bigl\langle\delta{\cal O}\bigr\rangle\con
  \enum
}
a number of times, where $\delta{\cal O}$ derives from
the transformation law 
\equation{
  \delta\psi_1=\dirac{5}(1-aD)\psi_1,
  \qquad
  \delta\psibar\vphantom{\psi}_1=\psibar\vphantom{\psi}_1\dirac{5}.
  \enum
}
As a result, a sum of connected 
expectation values of products of $m_rP_{st}$ and
$m_rS_{st}$ with $5$ or more factors is obtained,
all of which have only integrable
short-distance singularities in the continuum theory
(it should be noted here that
the constant field does not contribute to 
the operator product expansion in a connected correlation function
if only some coordinates are scaled to a common point).
The bottom line is then that the moments $C_n$ 
are finite in the continuum limit (if $\Nf\geq5$)
and that their values in the limit can, in principle, be determined
without recourse to any particular regularization of the theory.

\section 6

{\bf 6.}~For real-world QCD 
the condition that there should be at least $5$ flavours of quarks
is not a limitation, but it seems a bit strange
that the bottom quark must be involved in order 
to define the topological susceptibility in the continuum theory.
The constraint on the number of quark flavours can in fact be relaxed
by introducing a multiplet of valence quarks. 

In the extreme case of the pure gauge theory, 
for example, the fermion action on the lattice is then given by 
\equation{
  S_{\rm F}=a^4\sum_x\left\{\sum_{r=1}^{2N}\psibar_{r}(x)D_m\psi_r(x)+
  \sum_{k=1}^N\left|D_m\phi_k(x)\right|^2\right\},
  \enum
}
where $\psibar_r,\psi_r$ are the valence quark fields and $\phi_k$ 
the associated pseudo-fermion fields.
For positive quark masses $m$, 
a well-defined euclidean
field theory is obtained in this way
that is probably not reflection-positive, but 
in which power-counting arguments and the operator product expansion 
can be expected to apply,
since the locality of the theory is
preserved.

We may now again make use of the trace identities to rewrite the 
moments $C_n$ of the charge distribution 
in terms of correlation functions 
of the scalar and pseudo-scalar quark densities.
The discussion of the short-distance singularities is, however, slightly 
modified with respect to the case of full QCD, because
the pseudo-fermion fields allow for the construction of
flavour-singlet fields of dimension $2$. 
On the other hand, we can take advantage of the fact that 
there is no constraint on the number of valence 
quarks (the charge distribution is always the same), and
it is then easy to show that 
the expression
\equation{
  C_n=m^{6n}a^{24n-4}\sum_{x_1,\ldots,x_{6n-1}}
  \bigl\langle
  P_{31}(x_1)S_{12}(x_2)S_{23}(x_3)
  P_{64}(x_4)S_{45}(x_5)S_{56}(x_6)
  \ldots
  \noenum
  \next{1.5ex}
  \hskip5em
  \ldots P_{6n\kern1pt 6n-2}(x_{6n-2})S_{6n-2\kern1pt 6n-1}(x_{6n-1})
  S_{6n-1\kern1pt 6n}(0)  
  \bigr\rangle\con
  \enum
}
is finite in the continuum limit.

It may be worth mentioning here that 
we never had to refer to the chiral limit in this section
and that one is free to set the renormalized valence quark mass
to any fixed physical value.
This saved us from some difficult questions,
since the true asymptotic chiral behaviour of valence quarks is 
still not known.

Equation (6.2) can also be written in the more suggestive form
\equation{
   C_n=V^{-1}\bigl\langle \left(Q_{\rm F}\right)^{2n}
   \bigr\rangle\con,
   \enum
   \next{2ex}
   Q_{\rm F}\equiv -m^3a^6\sum_{x_1,x_2,x_3}
   \left\langle P_{31}(x_1)S_{12}(x_2)S_{23}(x_3)\right\rangle_{\rm F},
   \enum
}
where $V$ denotes the space-time volume.
In terms of the valence quark propagator, $Q_{\rm F}$ 
is simply equal to the triangle graph with
one pseudo-scalar and two scalar vertices
and vanishing external momenta. 
The important point is that these equations
apply independently of the chosen regularization,
provided the fields and the quark mass are renormalized
in accordance with the non-singlet chiral Ward identities.

\section 7

{\bf 7.}~Beyond the semi-classical regime, 
the universality of the moments of the topological charge distribution
(and thus of the distribution itself) 
derives from a combination of fundamental properties of QCD.
Asymptotic freedom was implicitly used, for example, when applying
the operator product expansion to show the
absence of non-integrable short-distance singularities in the final
formulae for the moments.
In the way the problem was approached here,
a key r\^ole was also played by the trace identities
(3.8) on the lattice and (4.2) in the continuum theory.
To some extent at least, these identities can be understood
as an algebraic reflection of 
the fact that the index of the Dirac operator
is a homotopy invariant.

If a lattice formulation of 
QCD that preserves chiral symmetry is used,
the naive definition of the topological susceptibility and
the higher moments coincides with the universal definition
in the continuum limit through the finite
expressions that were obtained in this paper.
This result is quite important, from both the conceptual
and the practical points of view,
and it also closes a gap
in the recent literature on the topological susceptibility
and the $\epsilon$-regime
(see ref.~[\ref{RandomMatrix}], for example, 
and references quoted there).
Evidently there is no reason to expect the same to be true  
if other formulations of lattice QCD and other
definitions of the topological charge density 
are considered. The moments can, however,
always be defined through
the universal formulae,
and some of these may actually be quite accessible to 
numerical simulations.

\vskip1ex
I wish to thank Leonardo Giusti for correspondence 
on the subject of this paper
and discussions on the derivation of the 
Witten--Veneziano formula.
Thanks also go to Peter Weisz for his comments on a first draft of the paper.

\beginbibliography

% Witten-Veneziano formula

\bibitem{Witten}
E. Witten, 
Nucl. Phys. B156 (1979) 269

\bibitem{Veneziano}
G. Veneziano,
Nucl. Phys. B159 (1979) 213; Phys. Lett. B95 (1980) 90

\bibitem{WVlattice}
L. Giusti, G. C. Rossi, M. Testa, G. Veneziano,
Nucl. Phys. B628 (2002) 234

\bibitem{Seiler}
E. Seiler,
Phys. Lett. B525 (2002) 355

% Reorganisation of the chiral expansion for small volumes and quark masses
% and systematic derivation of effective chiral Lagrangian in finite volume

\bibitem{GasserLeutwyler}
J. Gasser, H. Leutwyler,
Phys. Lett. B188 (1987) 477;
Nucl. Phys. B307 (1988) 763

% Clarification of the role of topology for spontaneous chiral symmetry
% breaking

\bibitem{LeutwylerSmilga}
H. Leutwyler, A. Smilga,
Phys. Rev. D46 (1992) 5607

% Non-topological charge densities

\bibitem{DiVecchiaEtAl}
P. DiVecchia, K. Fabricius, G. C. Rossi, G. Veneziano,
Nucl. Phys. B192 (1981) 392

% Classification of topological fields in the continuum theory

\bibitem{BrandtEtAl}
F. Brandt, N. Dragon, M. Kreuzer,
Phys. Lett. B231 (1989) 263;
Nucl. Phys. B332 (1990) 224 and 250

\bibitem{DuboisVioletteEtAl}
M. Dubois--Violette, M. Henneaux, M. Talon, C.--M. Viallet,
Phys. Lett. B267 (1991) 81;
{\it ibid.}\/ B289 (1992) 361

% Geometric constructions of the topological charge

\bibitem{TopA}
M. L\"uscher,
Commun. Math. Phys. 85 (1982) 39

\bibitem{TopB}
A. V. Phillips, D. A. Stone,
Commun. Math. Phys. 103 (1986) 599;
{\it ibid.}\/ 131 (1990) 255

% Exact chiral symmetry on the lattice

% Ginsparg-Wilson relation

\bibitem{GinspargWilson}
P. H. Ginsparg, K. G. Wilson,
Phys. Rev. D25 (1982) 2649

% Domain wall fermions

\bibitem{Kaplan}
D. B. Kaplan,
Phys. Lett. B288 (1992) 342;
Nucl. Phys. B (Proc. Suppl.) 30 (1993) 597

\bibitem{Shamir}
Y. Shamir,
Nucl. Phys. B406 (1993) 90

\bibitem{FurmanShamir}
V. Furman, Y. Shamir,
Nucl. Phys. B439 (1995) 54

% Classically perfect Dirac operator

\bibitem{Hasenfratz}
P. Hasenfratz,
Nucl. Phys. B (Proc. Suppl.) 63 (1998) 53;
Nucl. Phys. B525 (1998) 401

\bibitem{HLN}
P. Hasenfratz, V. Laliena, F. Niedermayer,
Phys. Lett. B427 (1998) 125

% Neuberger-Dirac operator

\bibitem{NeubergerDirac}
H. Neuberger,
Phys. Lett. B417 (1998) 141;
{\it ibid.}\/ B427 (1998) 353

% Exact chiral symmetry

\bibitem{ExactChSy}
M. L\"uscher,
Phys. Lett. B428 (1998) 342

% Locality of the Neuberger-Dirac operator, Chebyshev approximation

\bibitem{Locality}
P. Hern\'andez, K. Jansen, M. L\"uscher,
Nucl. Phys. B552 (1999) 363

\bibitem{BoulderReview}
F. Niedermayer,
Nucl. Phys. B (Proc. Suppl.) 73 (1999) 105

% Cancellation of power divergencies

\bibitem{GiustiRossiTesta} 
L. Giusti, G. C. Rossi, M. Testa,
hep-lat/0402027

% Proof of the index theorem using the heat equation

\bibitem{Gilkey}
P. B. Gilkey,
Invariance theory, the heat equation and the Atiyah--Singer index theorem,
2nd ed. (CRC Press, Boca Raton, 1995)

% Numerical study of the topological susceptibility

\bibitem{RandomMatrix}
L. Giusti, M. L\"uscher, P. Weisz, H. Wittig,
J. High Energy Phys. 0311 (2003) 023

\endbibliography

\bye